\documentclass[aps,prl,nopacs,twocolumn,twoside]{revtex4}
\usepackage{amsmath,amssymb,revsymb,epsfig}
\usepackage{graphicx}
\usepackage{theorem}

\newlength{\blank}
\newcommand{\ket}[1]{\left | #1 \right \rangle}
\newcommand{\bra}[1]{\left \langle #1 \right |}

\begin{document}

\title{Localization and its consequences\protect\\ for quantum walk algorithms and quantum communication}

\author{J.P.~Keating}
%%% jm delete 09 jun 06
%\email{j.p.keating@bristol.ac.uk}

\author{N.~Linden}
%%% jm delete 09 jun 06
%\email{n.linden@bristol.ac.uk}

\author{J.C.F.~Matthews}
%%% jm insertion 16 may 06 begin
%%% jm delete 09 jun 06
%\email{jonathan.matthews@bristol.ac.uk}
%%%% jm insertion 16 may 06 end
%%\email{j.matthews@bristol.ac.uk}

\author{A.~Winter}
%%% jm delete 09 jun 06
%\email{a.j.winter@bris.ac.uk}

\affiliation{Department of Mathematics, University of Bristol, Bristol BS8 1TW, United Kingdom}

\date{23rd June 2006}

\begin{abstract}
The exponential speed-up of quantum walks on certain graphs,
relative to classical particles diffusing on the same graph, is a
striking observation.  It has suggested the possibility of new
fast quantum algorithms.  We point out here that quantum mechanics
can also lead, through the phenomenon of localization, to
exponential suppression of motion on these graphs (even in the
absence of decoherence).  In fact, for physical embodiments of
graphs, this will be the generic behaviour. It also has
implications for proposals for using spin networks, including spin
chains, as quantum communication channels.
\end{abstract}

\maketitle

%%% jm removal 30 may 06 begin
%{\bf Introduction.}
%%% jm removal 30 may 06 end

The starting point for the field of quantum computation is the
realization that quantum algorithms may perform tasks more
efficiently than known classical algorithms for the same
problem~\cite{Feynman}. The first example of a complexity
separation was discovered  by Deutsch and Jozsa
\cite{DeutschJozsa92}. Perhaps the most celebrated example is
Shor's factoring algorithm \cite{Shor94}, which is exponentially
more efficient than any known classical algorithm.
%%% jm removal 30 may 06 begin
%; however unlike the Deutsch-Jozsa problem, it is not proven whether
%is a true exponential separation between the quantum and classical
%problems.
%%% jm removal 30 may 06 end

A third class of quantum algorithms stem from random walks,
%%% jm removal 01 jun 06 begin
%.  These
%%% jm removal 01 jun 06 end
%%% jm insertion 01 jun 06 begin
which
%%% jm insertion 01 jun 06 end
have proved a fruitful tool for finding classical algorithms
\cite{ClassicalRandomWalkAlgorithms1}. And this has led to
investigations of whether quantum random walks might offer
additional benefit for producing fast algorithms.  The first
realization was that quantum random walks in one dimension
propagate at a rate that is linear in time, compared to the
square-root rate for classical random walks
\cite{SquareRootSpeedUp}. This already suggests that quantum walks
may offer possibilities for at least polynomial speed-up of
algorithms.  Thus the subsequent discovery of quantum walks that
are exponentially faster than classical walks is particularly
striking and suggestive
\cite{ChildsFarhi,ExponentialSpeedUp,Kempe05}.

The clear message  is that quantum effects can enhance
computation.  In this paper we point out that the opposite is also
possible and that quantum effects can cause exponential
suppression of computations.

Our focus is on certain continuous time  walks (in contrast to the
discrete time ``coined'' walks mostly used in algorithmic
applications) which, when carried out on ideally perfect quantum
graphs, have exponentially faster propagation than a classical
random walk on the same graph. We will apply well-established
ideas from the theory of localization \cite{Anderson} to these
quantum walks to show that when the graphs have imperfections, as
they surely will  in any real physical implementation, the
propagation of quantum information is suppressed exponentially in
the amount of imperfection.

Our conclusion is that quantum walks on \emph{physical graphs} are
not likely to be useful for algorithmic purposes. It is worth
stressing at this point, however, that applications of quantum
walks in quantum algorithms are thought of as being simulated
walks in the memory of a quantum computer -- as is indeed always
the case in the classical applications where the underlying graphs
are typically exponentially large and hence would require
exponential resources to realize physically.
%%% insertion jm 16 may 06 begin
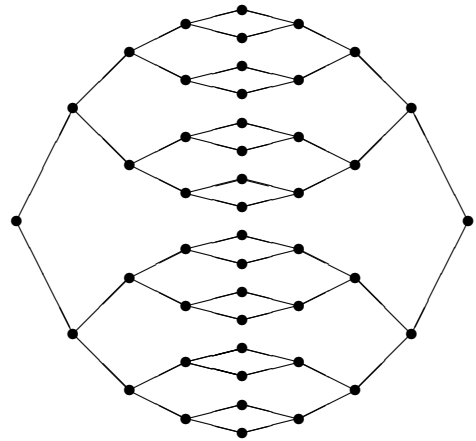
\begin{figure}[ht]
    \centering
    \setlength{\unitlength}{0.75cm}
    \begin{picture}(4,7.5)
%Left Half of graph
        \put(-2,3.75){\line(1,2){1}}
        \put(-2,3.75){\line(1,-2){1}}
        \put(-1,1.75){\line(1,-1){1}}
        \put(-1,1.75){\line(1,1){1}}
        \put(-1,5.75){\line(1,1){1}}
        \put(-1,5.75){\line(1,-1){1}}
        \put(0,6.75){\line(2,1){1}}
        \put(0,6.75){\line(2,-1){1}}
        \put(0,4.75){\line(2,1){1}}
        \put(0,4.75){\line(2,-1){1}}
        \put(0,0.75){\line(2,1){1}}
        \put(0,0.75){\line(2,-1){1}}
        \put(0,2.75){\line(2,1){1}}
        \put(0,2.75){\line(2,-1){1}}
        \put(1,7.25){\line(4,1){1}}
        \put(1,7.25){\line(4,-1){1}}
        \put(1,6.25){\line(4,1){1}}
        \put(1,6.25){\line(4,-1){1}}
        \put(1,5.25){\line(4,1){1}}
        \put(1,5.25){\line(4,-1){1}}
        \put(1,4.25){\line(4,1){1}}
        \put(1,4.25){\line(4,-1){1}}
        \put(1,1.25){\line(4,1){1}}
        \put(1,1.25){\line(4,-1){1}}
        \put(1,0.25){\line(4,1){1}}
        \put(1,0.25){\line(4,-1){1}}
        \put(1,1.25){\line(4,1){1}}
        \put(1,1.25){\line(4,-1){1}}
        \put(1,2.25){\line(4,1){1}}
        \put(1,2.25){\line(4,-1){1}}
        \put(1,3.25){\line(4,1){1}}
        \put(1,3.25){\line(4,-1){1}}
%Right Half of Graph
        \put(6,3.75){\line(-1,2){1}}
        \put(6,3.75){\line(-1,-2){1}}
        \put(5,1.75){\line(-1,-1){1}}
        \put(5,5.75){\line(-1,1){1}}
        \put(5,1.75){\line(-1,1){1}}
        \put(5,5.75){\line(-1,-1){1}}
        \put(4,6.75){\line(-2,1){1}}
        \put(4,6.75){\line(-2,-1){1}}
        \put(4,4.75){\line(-2,1){1}}
        \put(4,4.75){\line(-2,-1){1}}
        \put(4,0.75){\line(-2,1){1}}
        \put(4,0.75){\line(-2,-1){1}}
        \put(4,2.75){\line(-2,1){1}}
        \put(4,2.75){\line(-2,-1){1}}
        \put(3,7.25){\line(-4,1){1}}
        \put(3,7.25){\line(-4,-1){1}}
        \put(3,6.25){\line(-4,1){1}}
        \put(3,6.25){\line(-4,-1){1}}
        \put(3,5.25){\line(-4,1){1}}
        \put(3,5.25){\line(-4,-1){1}}
        \put(3,4.25){\line(-4,1){1}}
        \put(3,4.25){\line(-4,-1){1}}
        \put(3,0.25){\line(-4,1){1}}
        \put(3,0.25){\line(-4,-1){1}}
        \put(3,1.25){\line(-4,1){1}}
        \put(3,1.25){\line(-4,-1){1}}
        \put(3,2.25){\line(-4,1){1}}
        \put(3,2.25){\line(-4,-1){1}}
        \put(3,3.25){\line(-4,1){1}}
        \put(3,3.25){\line(-4,-1){1}}
%Nodes
        \put(6,3.75){\circle*{0.2}}
        \put(-2,3.75){\circle*{0.2}}
        \put(5,5.75){\circle*{0.2}}
        \put(-1,5.75){\circle*{0.2}}
        \put(5,1.75){\circle*{0.2}}
        \put(-1,1.75){\circle*{0.2}}
        \put(4,6.75){\circle*{0.2}}
        \put(4,0.75){\circle*{0.2}}
        \put(0,6.75){\circle*{0.2}}
        \put(0,0.75){\circle*{0.2}}
        \put(4,4.75){\circle*{0.2}}
        \put(0,4.75){\circle*{0.2}}
        \put(4,2.75){\circle*{0.2}}
        \put(0,2.75){\circle*{0.2}}
        \put(3,7.25){\circle*{0.2}}
        \put(1,7.25){\circle*{0.2}}
        \put(3,0.25){\circle*{0.2}}
        \put(1,0.25){\circle*{0.2}}
        \put(3,6.25){\circle*{0.2}}
        \put(3,1.25){\circle*{0.2}}
        \put(1,6.25){\circle*{0.2}}
        \put(1,1.25){\circle*{0.2}}
        \put(3,5.25){\circle*{0.2}}
        \put(3,2.25){\circle*{0.2}}
        \put(1,5.25){\circle*{0.2}}
        \put(1,2.25){\circle*{0.2}}
        \put(3,4.25){\circle*{0.2}}
        \put(3,3.25){\circle*{0.2}}
        \put(1,4.25){\circle*{0.2}}
        \put(1,3.25){\circle*{0.2}}
        \put(2,7.5){\circle*{0.2}}
        \put(2,7){\circle*{0.2}}
        \put(2,6.5){\circle*{0.2}}
        \put(2,6){\circle*{0.2}}
        \put(2,5.5){\circle*{0.2}}
        \put(2,5){\circle*{0.2}}
        \put(2,4.5){\circle*{0.2}}
        \put(2,4){\circle*{0.2}}
        \put(2,3.5){\circle*{0.2}}
        \put(2,3){\circle*{0.2}}
        \put(2,2.5){\circle*{0.2}}
        \put(2,2){\circle*{0.2}}
        \put(2,1.5){\circle*{0.2}}
        \put(2,1){\circle*{0.2}}
        \put(2,0.5){\circle*{0.2}}
        \put(2,0){\circle*{0.2}}
    \end{picture}
    \caption{\footnotesize{The Graph ${\cal G}_4$ \cite{ChildsFarhi}.}}
    \label{figure1}
\end{figure}
%%% insertion jm 16 may 06 end
%%% jm insertion 16 may 06 begin

%%% jm removal 01 jun 06 begin
%%%% jm insertion 16 may 06 end
%The particular case we will consider is the graph first studied in
%\cite{ChildsFarhi}.  It consists of two branching trees joined
%back-to-back; it is illustrated in Fig
%%%% jm insertion 16 may 06 begin
%\ref{figure1}.
%%%% jm insertion 16 may 06 end
%This graph is called ${\cal G}_n$ where $2n+1$ is the number of
%vertical columns of nodes.
%%% jm removal 01 jun 06 end
%%% jm insertion 01 jun 06 begin
We describe in detail a particular example of propagation on a
graph which, in the absence of imperfections, has the property
that quantum evolution is exponentially faster than classical
evolution; the effect of localization is particularly stark here.
Later we point out the implications for quantum information
processing tasks on other networks.

The particular case we will consider is the graph ${\cal G}_n$,
first studied in \cite{ChildsFarhi}. It consists of two joined
branching trees  with $2n+1$ vertical columns of nodes; ${\cal
G}_4$ is illustrated in Figure \ref{figure1}.
%%% jm insertion 01 jun 06 end
The question of interest is how rapidly a particle starting at the
left-most node reaches the right-most node.  Let us first describe
the idea behind the exponential separation between classical
diffusion and quantum motion on this graph (when the graph is
perfect).  Our treatment follows that in \cite{ChildsFarhi,Kempe} closely.

The model of classical diffusion is that the particle is equally
likely to move from the node where it finds itself to any of the
nodes to which it is linked.
%%% jm removal 09 jun 06 begin
%Explicitly we may take the motion to arise from the following
%Hamiltonian.
%%% jm removal 09 jun 06 end
%%% jm insertion 09 jun 06 begin
Explicitly we may take the motion to arise from the Markov
transition matrix
\begin{eqnarray}
 \label{ClassicalHamiltonian1} M_{i,j}=\left\{
\begin{array}{cl}
   -\gamma &  \ i \neq j \mbox{ if nodes }i\mbox{ and }j\mbox{ connected} \\
         0 &  \ i \neq j \mbox{ if nodes }i\mbox{ and }j\mbox{ not connected} \\
d_i \gamma &  \ i=j
\end{array} \right.
\end{eqnarray}
where $\gamma$ is the constant jumping rate between two connecting
vertices, and $d_i$ is the number of edges incident to the vertex
$i$. The probability of being at node $i$ at time $t$ is governed by
%
%%% jm removal 10 jun 06 begin
%\begin{eqnarray}
%\frac{d p_i(t)}{dt}=-\sum_j H_{i,j}p_j(t).
%\end{eqnarray}
%%% jm removal 10 jun 06 end
%
%%% jm insertion 10 jun 06 begin
\begin{eqnarray}
\frac{d p_i(t)}{dt}=-\sum_j M_{i,j}p_j(t).
\end{eqnarray}
%%% jm insertion 10 jun 06 end

A classical particle finding itself in the middle of the graph
(not at the left- or right-most node or one of the nodes precisely
in the centre) is twice as likely to move towards the centre as
away from it. Thus a classical particle  starting at the left-most
node will diffuse rapidly to the centre of the graph but then it
will diffuse exponentially slowly from the centre of the graph to
the rightmost node.
%%% jm removal 31 may 06 begin
%For this graph, the Hamiltonian is
%%% jm insertion 16 may 06 begin
%\begin{eqnarray}
%    H_{\tilde{i},\tilde{j}}=\left\{
%        \begin{array}{cl}
%            1/2   &  \tilde{i} = \tilde{j} \pm 1 \mbox{ and } \tilde{i}=n+1 \\
%            2/3   &  \tilde{i} = \tilde{j} + 1 \mbox{ and } 1 < \tilde{i} < n \\
%            2/3   &  \tilde{i} = \tilde{j} - 1 \mbox{ and } n < \tilde{i} < 2n + 1 \\
%            1/3   &  \tilde{i} = \tilde{j} - 1 \mbox{ and } 1 < \tilde{i} < n \\
%            1/3   &  \tilde{i} = \tilde{j} + 1 \mbox{ and } n < \tilde{i} < 2n + 1 \\
%            1     &  \tilde{i} = \tilde{j} + 1 \mbox{ and } \tilde{i}= n \\
%            1     &  \tilde{i} = \tilde{j} + 1 \mbox{ and } \tilde{i}= 1 \\
%            0     &  \mbox{ otherwise }
%\end{array} \right.
% \label{ReducedClassicalHamiltonian}
%{\rm reduced\ classical\ Hamiltonian}
%\end{eqnarray}
%%% jm insertion 16 may 06 end
%%% jm removal 31 may 06 end

We now analyze the quantum walk on ${\cal G}_n$ \cite{ChildsFarhi}
starting in the state corresponding to the left-most node and
evolving with the Hamiltonian given by
\begin{eqnarray}
    \left\langle a | H | b \right\rangle = M_{a,b},
 \label{QuantumHamiltonian1}
%{\rm quantum\ Hamiltonian}
\end{eqnarray}
where $\left|b\right\rangle$ represents the state of a particle at
node $b$.
%%% jm insertion 09 jun 06 end

The key observation is that
%%% jm removal 30 may 06 begin
%the
%%% jm removal 30 may 06 end
although there are exponentially many nodes, and a Hilbert space of
exponential dimension, in fact the system only evolves in a
$(2n+1)$-dimensional subspace. This subspace is spanned by states
$|\tilde{j}\rangle$ (where $0 \le j \le 2n$), the uniform
superposition over all vertices in column $j$, that is,
\begin{eqnarray}
  |\tilde{j}\rangle
    =2^{-\min[j,2n-j]/2} \sum_{a \in\ {\rm column}\ {j}} |a\rangle
.
\end{eqnarray}

\begin{figure}[ht]
    \centering
    \setlength{\unitlength}{0.9cm}
    \begin{picture}(9,1)
        \put(0.55,0.5){\line(1,0){8}}
        \put(0.55,0.5){\circle*{0.2}}
        \put(1.55,0.5){\circle*{0.2}}
        \put(2.55,0.5){\circle*{0.2}}
        \put(3.55,0.5){\circle*{0.2}}
        \put(4.55,0.5){\circle*{0.2}}
        \put(5.55,0.5){\circle*{0.2}}
        \put(6.55,0.5){\circle*{0.2}}
        \put(7.55,0.5){\circle*{0.2}}
        \put(8.55,0.5){\circle*{0.2}}
        \put(0.6,0.75){{\scriptsize$-\sqrt{2}$}}
        \put(1.6,0.75){{\scriptsize$-\sqrt{2}$}}
        \put(2.6,0.75){{\scriptsize$-\sqrt{2}$}}
        \put(3.6,0.75){{\scriptsize$-\sqrt{2}$}}
        \put(4.6,0.75){{\scriptsize$-\sqrt{2}$}}
        \put(5.6,0.75){{\scriptsize$-\sqrt{2}$}}
        \put(6.6,0.75){{\scriptsize$-\sqrt{2}$}}
        \put(7.6,0.75){{\scriptsize$-\sqrt{2}$}}
        \put(0.475,0){{\scriptsize$2$}}
        \put(1.475,0){{\scriptsize$3$}}
        \put(2.475,0){{\scriptsize$3$}}
        \put(3.475,0){{\scriptsize$3$}}
        \put(4.475,0){{\scriptsize$2$}}
        \put(5.475,0){{\scriptsize$3$}}
        \put(6.475,0){{\scriptsize$3$}}
        \put(7.475,0){{\scriptsize$3$}}
        \put(8.475,0){{\scriptsize$2$}}
    \end{picture}
%%% jm removal 01 jun 06 begin
%    \caption{\footnotesize{The Graph ${\cal G}_4$ is reduced to a quantum random walk on a line with $9$ vertices.}}
%%% jm removal 01 jun 06 end
%%% jm insertion 01 jun 06 begin
    \caption{\footnotesize{The elements of the Hamiltonian $H/{\gamma}$ for ${\cal G}_4$ when reduced to a walk on a line \cite{ChildsFarhi}.}}
%%% jm insertion 01 jun 06 end
    \label{figure3}
\end{figure}
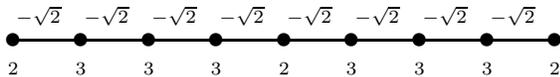

In this basis, the non-zero matrix elements of $H$ are
\begin{eqnarray}\label{QuantumHamiltonianOnLine}
  \langle \tilde{j}|H|\tilde{j} \pm 1\rangle &=& -\sqrt{2}\gamma \nonumber\\
 \langle \tilde{j}|H|\tilde{j}\rangle &=& \left\{ \begin{array}{ll}
    2\gamma & {j}=0,n,2n \\
    3\gamma & {\rm otherwise,} \\
    \end{array} \right.
\end{eqnarray}
which is depicted in Figure
%%% jm insertion 16 may 06 begin
\ref{figure3}
%%% jm insertion 16 may 06 end
as a quantum random walk on a line with $2n+1$ vertices.

%%% jm insertion 16 may 06 begin
%%% jm insertion 16 may 06 end

In particular it is seen that (away from the left-most or right-most
%%% jm insertion 30 may 06 begin
vertices%%% jm insertion 30 may 06 end
) the quantum particle has equal amplitude to move to the left as to
the right, and in particular it is not exponentially suppressed from
moving right once it has reached the centre.
(Observe that the central node, coming from the centre of the graph
${\cal G}_n$, has a special status, but that will us not concern here.)

We now turn to the main point of this paper.  If we were to try and
build this graph physically and get a quantum particle to evolve on
it, it would be virtually impossible to do
%%% jm remove 09 jun 06 :"it"
%%% jm insert 09 jun 06 begin
so
%%% jm insert 09 jun 06 end
perfectly.  In particular, for example, the distances between the
nodes will inevitably vary slightly from edge to edge.  Our main
observation will be that the theory of Anderson localization
\cite{Anderson} implies that this variability will lead to
suppression of the quantum evolution so that in fact a quantum
particle starting at the left of the graph can only travel a
distance to the right proportional to an inverse power of the
degree of variability. The probability of arriving at a point
beyond this \lq\lq localization length\rq\rq\ is exponentially
small in the distance from the starting point. Thus if the
right-most node is further away than the localization length, the
particle is effectively prevented from reaching it. We emphasize
that localization is very much a quantum (or more strictly a wave)
phenomenon; it does not affect classical particles diffusing.
Indeed, Anderson localization is often characterized as the
quantum suppression of classical diffusion. We feel it is striking
that this is a case where quantum effects are suppressing rather
than enhancing possibilities for information processing.

The imperfections in the quantum evolution we consider are within
the graph itself; namely the Hamiltonian varies a little from node
to node. But the evolution is unitary.  In particular, we are not
concerned with
%%% jm insertion 30 may 06 begin
any
%%% jm insertopn 30 may 06 end
interaction of the quantum system with the environment, in other
words we assume that there is no decoherence
\cite{DecoherenceInGraphs}. In fact we will consider
%%% jm insertion 30 may 06 begin
a
%%% jm insertion 30 may 06 end
rather
%%% jm removal 30 may 06 begin
% a
%%% jm removal 30 may 06 end
weak model of imperfection for our graph.  In the original
$(2^{n+1}+2^n-2)$-node graph, one expects that the interaction
connecting each pair of nodes will vary slightly from edge to edge.
This breaking of the symmetry of the original system would have a
%%% jm removal 30 may 06 begin
%one
%%% jm removal 30 may 06 end
substantial effect on the system, namely that the evolution would
now no longer proceed simply through the states $\ket{\tilde{j}}$,
but leak into the entire Hilbert space.  Our model of the
variability of the Hamiltonian is less severe than this: we assume
that the evolution still proceeds through the states
$\ket{\tilde{j}}$, but that this effective walk on the line is
subject to  a Hamiltonian that varies from node to node on the
line. It is to be expected that the more general situation in
which variability is allowed for all nodes/edges can only further
suppress the quantum evolution.  Thus our system is still
considered to walk on a one-dimensional graph, but now  our
Hamiltonian is
%
%%% jm removal 10 jun 06 begin
%\begin{eqnarray}\label{QuantumHamiltonian2}
%  \langle \tilde{j}|H|\tilde{j}\pm 1\rangle &=& -\sqrt{2}\gamma +\epsilon_{j} \nonumber \\
% \langle \tilde{j}|H|\tilde{j}\rangle &=& \left\{ \begin{array}{ll}
%    2\gamma & {j}=0,n,2n \\
%    3\gamma & {\rm otherwise,} \\
%    \end{array} \right.
%\end{eqnarray}
%%% jm removal 10 jun 06 end
%
%%% jm insertion 10 june 06 begin
\begin{eqnarray}\label{QuantumHamiltonian2}
  H_\epsilon = H + \sum_j \epsilon_j \ket{j}\!\bra{j},
\end{eqnarray}
%%% jm insertoon 10 june 06 end
%
where $H$ is the Hamiltonian (\ref{QuantumHamiltonianOnLine}). The
variability is introduced via the parameters $\epsilon_{j}$. In
the first instance  the parameters $\epsilon_{j}$ will be taken
independently from a Cauchy distribution with parameter $\delta$:
the density is
%%% jm insertion 17 may 06 begin
\begin{eqnarray}
P\left(\epsilon\right) = \frac{1}{\pi}\frac{\delta}{\epsilon^2 + \delta^2}.
\end{eqnarray}
%%% jm insertion 17 may 06 end

A particular reason for doing this is that the Cauchy distribution
is most amenable to analytic treatment.  Later we will point out
that taking the $\epsilon_{j}$ to be chosen from other
distributions makes no qualitative difference to the conclusions
(although it does
%%% jm removal 30 may 06 begin
%seem
%%% jm removal 30 may 06 end
lead to some interesting quantitative differences). It will also
be clear that we could encode variability in the Hamiltonian in
various other ways than (\ref{QuantumHamiltonian2}); for example
the off-diagonal terms could have random components. Many studies
in the localization literature \cite{DerridaGardner84,Slevin88}
show that these possibilities  do not make a qualitative change to
our conclusion.

\begin{figure}
    \centering
    \includegraphics[width=8cm]{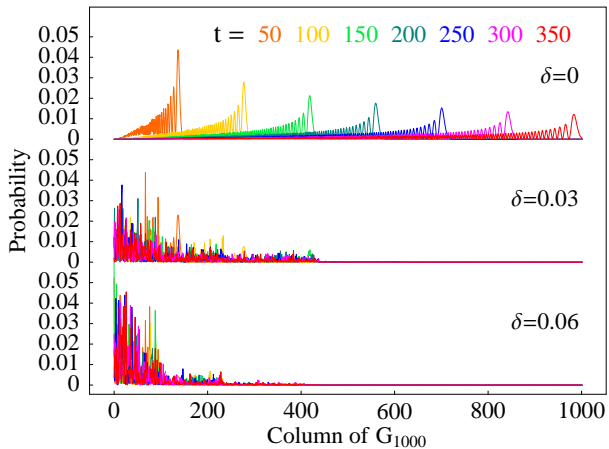}
    \caption{\footnotesize{(Colour online) Propagation within the left side of
       ${\cal G}_{1000}$ at times $t$ according to; (top) the perfect Hamiltonian
       eq.~(\ref{QuantumHamiltonian1}), and the disordered Hamiltonian
       eq.~(\ref{QuantumHamiltonian2}) with (middle) disorder $\delta=0.03$,
       and (bottom) disorder $\delta=0.06$.}}
    \label{figure4}
\end{figure}

All our conclusions are
%%% jm removal 30 may 06 begin
%simply
%%% jm removal 30 may 06 end
encapsulated in Figure \ref{figure4}.
%%% jm insertion 09 jun 06 end
The first plot shows the flow of intensity with time for motion
along left half of the columns of ${\cal G}_n$ ($0<j<n$)
%%% jm insertion 09 jun 06 end
according to the Hamiltonian (\ref{QuantumHamiltonianOnLine}) -
the perfect case. The signal clearly travels smoothly (and indeed
at constant speed) along the chain. The remaining two plots show
what happens when the Hamiltonian is now
(\ref{QuantumHamiltonian2}) with the $\epsilon_{j}$ taken from the
Cauchy distribution with increasing amounts of variability,
$\delta$.
%
%%% jm insertion 09 jun 06 begin
In all cases, we numerically computed the probability
$\left|\left\langle{\tilde{j}}|\psi\left(t\right)\right\rangle\right|^2$
of a particle being in the state $\left|{\tilde{j}}\right\rangle$
for seven times $t$, where
$\left|\psi\left(t\right)\right\rangle=\exp\left(-itH/\hbar\right)\left|\tilde{0}\right\rangle$,
$\gamma=1$, $\hbar=1$, and $n=1000$.
%

%%% jm insertion 09 jun 06 end

When $\delta$ is non-zero, the quantum wave packet travels a
certain distance along the graph, but is then seen to stop. The
distance that the packet gets to is reduced as $\delta$ increases
(in fact, as we shall see the distance is inversely proportional
to $\delta$).

This behaviour is a classic example of Anderson localization
\cite{Anderson}.
%%% jm removal 09 jun 06 begin
%It arises since the eigenstates of the Hamiltonian
%(\ref{QuantumHamiltonian2}), rather than being extended in space as
%in the case (\ref{QuantumHamiltonian1}), are exponentially localized
%in space.  The modulus of the amplitudes of $E_\alpha({\tilde j})$
%of each eigenstate
%\begin{eqnarray}
%\ket{E_\alpha}=\sum_\alpha E_\alpha({\tilde j})\ket {\tilde j}
%\end{eqnarray}
% is bounded by a function
%of the form
%%%% jm remove 17 may 06 begin
%%$e^{-\lambda_{\alpha} {\tilde j}}$;
%%%% jm remove 17 may 06 end
%%%% jm insertion 17 may 06 begin
%$e^{-\left|x_{\alpha} - {\tilde j}\right|/l_{\alpha}}$;
%%%% jm insertion 17 may 06 end
%%%% jm remove 17 may 06 begin
%%$\lambda_{\alpha}$
%%%% jm remove 17 may 06 end
%%%% jm insertion 17 may begin
%$l_{\alpha}$
%%%% jm inssertion 17 may 06 end
%is the localization length
%%%% jm insertion 17 may 06 begin
%and $x_\alpha$ is the centre
%%%% jm insertion 17 may 06 end
%of the eigenstate $\ket{E_\alpha}$.
%%% jm removal 09 jun 06 end
%
%%% jm insertion 09 jun 06 begin
It arises since the eigenstates, $\left|E_{\alpha}\right\rangle$,
of the Hamiltonian (\ref{QuantumHamiltonian2}) are exponentially
localized in space, rather than being extended into space as in
the case (\ref{QuantumHamiltonian1}). The modulus of the amplitude
of each eigenstate is bounded by an exponential function
\begin{eqnarray}
\left|\left\langle E_{\alpha}|{\tilde  j}\right\rangle\right|<
N_\alpha e^{-\left|{\tilde{j}}^{0}_{\alpha}-{\tilde
j}\right|/{l_{\alpha}}},
\end{eqnarray}
where $l_{\alpha}$ is the localization length and $N_\alpha$ is
a normalization constant. It is simple to see why this behaviour
of the eigenstates causes the time evolution witnessed in Figure
%%% jm insertion 17 may 06 begin
\ref{figure4}.
%%% jm insertion 17 may 06 end
The state of the system at time $t$ may be written as
\begin{eqnarray}
\ket{\psi(t)} = \sum_\alpha \psi_\alpha
e^{-itE_{\alpha}/\hbar}\ket{E_\alpha},
\end{eqnarray}
where $\psi_\alpha = \langle {E_\alpha} \ket{\psi(0)}$ are the
amplitudes of the initial state
%
%%% jm insertion 10 jun 06 begin
and $E_{\alpha}$ are the energy levels.
%%% jm insertion 01 jun 06 end
%
%%% jm remove 09 jun 06 begin
%in the basis $\ket{E_\alpha}$
%%% jm remove 09 jun 06 end
Now consider a quantum state initially concentrated at some point.
This initial state will only have substantial amplitude in energy
eigenstates which are localized in regions close to that initial
point. As the state evolves with time, the phase of these amplitudes
can change, but not the modulus of the amplitudes. Thus for all time
the state is expanded in terms of states which have small amplitude
far from the initial point.

For the particular model of Hamiltonian variability
(\ref{QuantumHamiltonian2}), we can explicitly describe the
behaviour of the localization length of eigenstates as $\delta$
varies, treating as negligible the influence of the anomalous end
and centre points of the graph. Namely, it has then been shown
\cite{LloydThouless} that
\begin{equation}
  \cosh{\frac{1}{l_{\alpha}}}
     = \frac{\sqrt{\left(\!\sqrt{8}\gamma \!+\! \hat E_{\alpha}\right)^2 \!\!+\! \delta^2}
             \!+\! \sqrt{\left(\!\sqrt{8}\gamma \!-\! \hat E_{\alpha}\right)^2 \!\!+\! \delta^2}}
            {\sqrt{32}\gamma},
  \label{cosh}
\end{equation}
where $\hat E_\alpha=E_\alpha-3\gamma$.   It is not difficult to
check that the largest localization length occurs for
$E_{\alpha}=3\gamma$, and for small $\delta$ (i.e. $\delta \ll
\sqrt{8}\gamma$) it becomes $l_{\rm max} \sim
\sqrt{8}\gamma/\delta$. Since the Hamiltonian has a random
component, it will almost surely not have an eigenvalue equal to
$3\gamma$; but the localization length for $E_\alpha=3\gamma$ is
an upper bound for the localization length of any eigenstate. And
hence this gives an estimate of the furthest distance a signal
starting at a given point will get to.
%%% jm remove 17 may 06 begin
%The figures xxx entirely
%%% jm remove 17 may 06 end
%%% jm insertion 17 may 06 begin
The plots in Figure \ref{figure4}
%%% jm insertion 17 may 06 end
bear this out.

%{\bf Conclusions.}
This then is the main conclusion of our investigation.  A
particularly quantum effect,  Anderson localization, causes a
particle undergoing a quantum walk in the graph to be effectively
stopped from propagating from left to right.  More precisely,
consider a fixed value of the degree of variability $\delta$, and a
series of graphs ${\cal G}_n$ with $n$ increasing. Once the length
of the graph $n$ is greater than about the maximum localization
length $\sqrt{8}\gamma/\delta$, the probability of the particle
reaching the right-most node, having started at the left-most node
is exponentially small in $n$.

This is reminiscent of the behaviour of the classical random walk
on the same graph.  However it is fundamental that the exponents
arise from quite different sources.  In particular we note that in
the original problem of  motion on the graph ${\cal G}_n$, a
classical particle diffusing from left to right travels rapidly
from the left most node to the centre, and the exponential time it
takes to get from left to right arises from the fact that it is
difficult to get from the centre of the graph to the right-most
node.  In the case of a quantum particle on an imperfect graph,
localization means that it can be exponentially difficult even to
get from the left-most node to the centre of the graph.

It is important to point out that localization is not a consequence
of our choice of Cauchy distributed disorder.
%%% jm removal 09 jun 06 begin
%Therefore
%%% jm removal 09 jun 06 end
It is a generic property of disordered systems, in one dimension,
occurring for any distribution. For other choices (e.g Gaussian or
uniform on an interval), the qualitative features are the same as
for the Cauchy case, but the quantitative details may differ; in
particular, if $P(\epsilon)$ has a second moment, the localization
length typically scales like $\delta^{-2}$ as $\delta \rightarrow
0$, rather than $\delta^{-1}$~\cite{DerridaGardner84}.

The graph ${\cal G}_n$ is of particular interest because of its
relationship to quantum algorithms and the exponential separation
between between classical and quantum propagation.  However our
results also have implications for other quantum information
processing tasks.  For example, there has been considerable
interest recently in using spin chains and other networks for
propagation quantum information \cite{spinchains}.  The message of
this paper is that localization will present a fundamental
difficulty for these proposals even in the absence of external
interactions with the system. (See also~\cite{deChiara}.)

To end this discussion, we emphasise two differences of the model
considered here to noisy quantum computation. First, noise in
quantum computers is usually modelled as sufficiently independent
stochastic variations in the dynamics, in particular on a single
memory qubit they vary with time; here, we have randomness in the
description of the system itself, i.e.~the Hamiltonian, but on the
other hand the localization effect remains even if the particular
random Hamiltonian is known. Second, we consider here free
evolution of the quantum state on a spatially extended system;
this means that any comparison with quantum computers should not
be with the circuit model (for which also techniques for
fault-tolerant computation exist \cite{fault-tolerant}), but rather
with computation in a closed system, like quantum cellular
automata \cite{qca}. It is not known whether fault-tolerant
techniques for universal quantum cellular automata in one
dimension apply to the type of error considered in localization:
fault-tolerant computing usually has to assume a degree of
independence of errors, both spatially and in time, and although
able to cope with certain correlations, the persistent
``failure'' of an interaction in always the same way throughout
the evolution seems to present a new challenge.

Also, it should be pointed out that localization is known to occur
for a single spin excitation moving in spin lattices, and it may
be expected for small numbers of excitations; however, universal
quantum computation happens in the regime of many excitations,
where it is not known if localization presents an obstacle for
information propagation.

We remark finally  that the systems we have considered here are
different to stroboscopic ($\delta$-kicked) models (e.g. those in
\cite{BuerschaperBoness}) which are closely related to the kicked
rotor, where localization occurs for quantum chaotic reasons
\cite{Fishman82}, and its consequences have been extensively
explored by Shepelyansky and co-workers (see for example
\cite{Shepelyansky01}).

\begin{acknowledgments}
  NL, JM and AW
  thank the U.K.~EPSRC for support through the IRC in QIP, as well
  as the EC for support through the QAP project. JPK is
  supported by an EPSRC Senior Research Fellowship;
  AW additionally through a University of Bristol
  Research Fellowship.
\end{acknowledgments}

\end{document}